\documentclass[3p,times,procedia]{elsarticle}
\usepackage{nupha_ecrc}
\usepackage{wrapfig}
\usepackage{lineno}
\graphicspath{{./img/}}

\volume{00}

\firstpage{1}

\journalname{Nuclear Physics A}

\runauth{}

\jid{nupha}

\jnltitlelogo{Nuclear Physics A}

\usepackage{amssymb}

\usepackage[figuresright]{rotating}

\begin{document}

\begin{frontmatter}



\dochead{XXVIIIth International Conference on Ultrarelativistic Nucleus-Nucleus Collisions\\ (Quark Matter 2019)}

\title{Jet shapes and fragmentation functions in Au$+$Au collisions at $\sqrt{s_{\mathrm{NN}}}=200$ GeV in STAR}


\author{Saehanseul Oh (for the STAR Collaboration)}

\address{Yale University, New Haven, Connecticut 06520}

\begin{abstract}
The STAR Collaboration reports measurements of differential jet shapes and semi-inclusive jet fragmentation functions in Au+Au collisions at $\sqrt{s_{\mathrm{NN}}}=$ 200 GeV with the STAR detector at RHIC. 
Jet shapes, which represent the radial distribution of momentum carried by constituents, are measured differentially for (1) the  charged particles transverse momentum and (2) the jet azimuthal angle relative to the second-order event plane. 
Based on the semi-inclusive population of jets recoiling from a high transverse momentum trigger hadron, jet fragmentation functions in 40-60\% central heavy-ion collisions are measured, and compared to those in PYTHIA simulations for $pp$ collisions. 
\end{abstract}


\end{frontmatter}



\section{Introduction}
\label{sec:intro}
Jet quenching, which refers to the interaction of a jet shower with Quark-Gluon Plasma (QGP) generated in relativistic heavy-ion collisions, is one of the key signals to probe the existence and properties of QGP (a recent review of jet studies in relativistic heavy ion physics is given in \cite{Connors:2017ptx}).   
In particular, the modification of jet substructure in heavy-ion collisions with respect to the vacuum reference has been investigated at the LHC with various observables, such as jet shapes, fragmentation functions, and shared momentum fraction. 
With increased data sample sizes during the recent RHIC runs and advanced techniques in handling background jets~\cite{Adamczyk:2016fqm, Adamczyk:2017yhe}, such measurements have become feasible at RHIC energies with the STAR detector. 
In these proceedings, we present two jet substructure measurements in heavy-ion collisions in STAR: (1) differential jet shapes and (2) semi-inclusive jet fragmentation functions. 

Both measurements commonly utilize Au+Au collisions at $\sqrt{s_{\mathrm{NN}}}=$ 200 GeV, collected in 2014 by the STAR experiment~\cite{Ackermann:2002ad}.
Events containing a high energy  Barrel Electromagnetic Calorimeter (BEMC) tower ($E_{\mathrm{T}} >$ 4.3 GeV) are used for signal jets, and minimum-bias events are used to estimate the effects of background via a mixed-event technique~\cite{Adamczyk:2017yhe}. 
Jets are reconstructed with the anti-$k_{\mathrm{T}}$ sequential jet clustering algorithm from the FastJet package~\cite{Cacciari:2011ma} with a radius parameter $R =$ 0.4. 
While the fragmentation function measurement uses only charged tracks reconstructed in the Time Projection Chamber (TPC) to obtain charged jets, the jet shape measurement is based on full jets which additionally include energy deposited in the BEMC towers as neutral jet constituents. 
Mixed events are formulated for distinguished classes of events based on their multiplicity, primary vertex position along the beam direction, and the second-order event-plane angle ($\Psi_{\mathrm{EP}}$), and used in both measurements to subtract background contributions.

\section{Differential jet shapes}
\label{sec:JetShape}
The jet shape function, $\rho(\Delta r)$, provides information about the radial distribution of the momentum carried by the jet constituents~\cite{Chatrchyan:2013kwa}.
The jet shape function is defined as 
\begin{eqnarray}
\rho(\Delta r) = \frac{1}{\delta r}\frac{1}{N_{\mathrm{jet}}}\sum_{\mathrm{jet}}\frac{\sum_{\mathrm{track}\in(r_{\mathrm{a}}, r_{\mathrm{b}})} p_{\mathrm{T,track}}}{p_{\mathrm{T,jet}}} \, \mathrm{,}
\end{eqnarray}
where $\Delta r = \sqrt{(\varphi_{\mathrm{track}} - \varphi_{\mathrm{jet}})^2 + (\eta_{\mathrm{track}} - \eta_{\mathrm{jet}})^2}$, $r_{\mathrm{a}} = \Delta r-\delta r/2$, $r_{\mathrm{b}} = \Delta r + \delta r /2$, and $\delta r$ is the radial annulus bin size. 
In order to suppress the effects of background fluctuations and combinatorial jets not originating from an initial hard scatter, jets are first reconstructed with tracks and BEMC towers with $p_{\mathrm{T}} > 2.0$ GeV/$c$ and $E_{\mathrm{T}} > 2.0$ GeV, respectively. 
These jets have been previously called ``hard-core'' jets, and no event-by-event background energy subtraction is applied for the $p_{\mathrm{T,jet}}$ calculation~\cite{Adamczyk:2016fqm}. 
Then $\rho(\Delta r)$ is estimated by associating these jets with charged tracks with $1.0 < p_{\mathrm{T,track}} < 30.0$ GeV/$c$ and $r<R$. 
For these proceedings, results with only the highest-$p_{\mathrm{T}}$ jets in each event (leading jets) are presented. 

\begin{figure}
	\centering
	\includegraphics[width=0.76\textwidth]{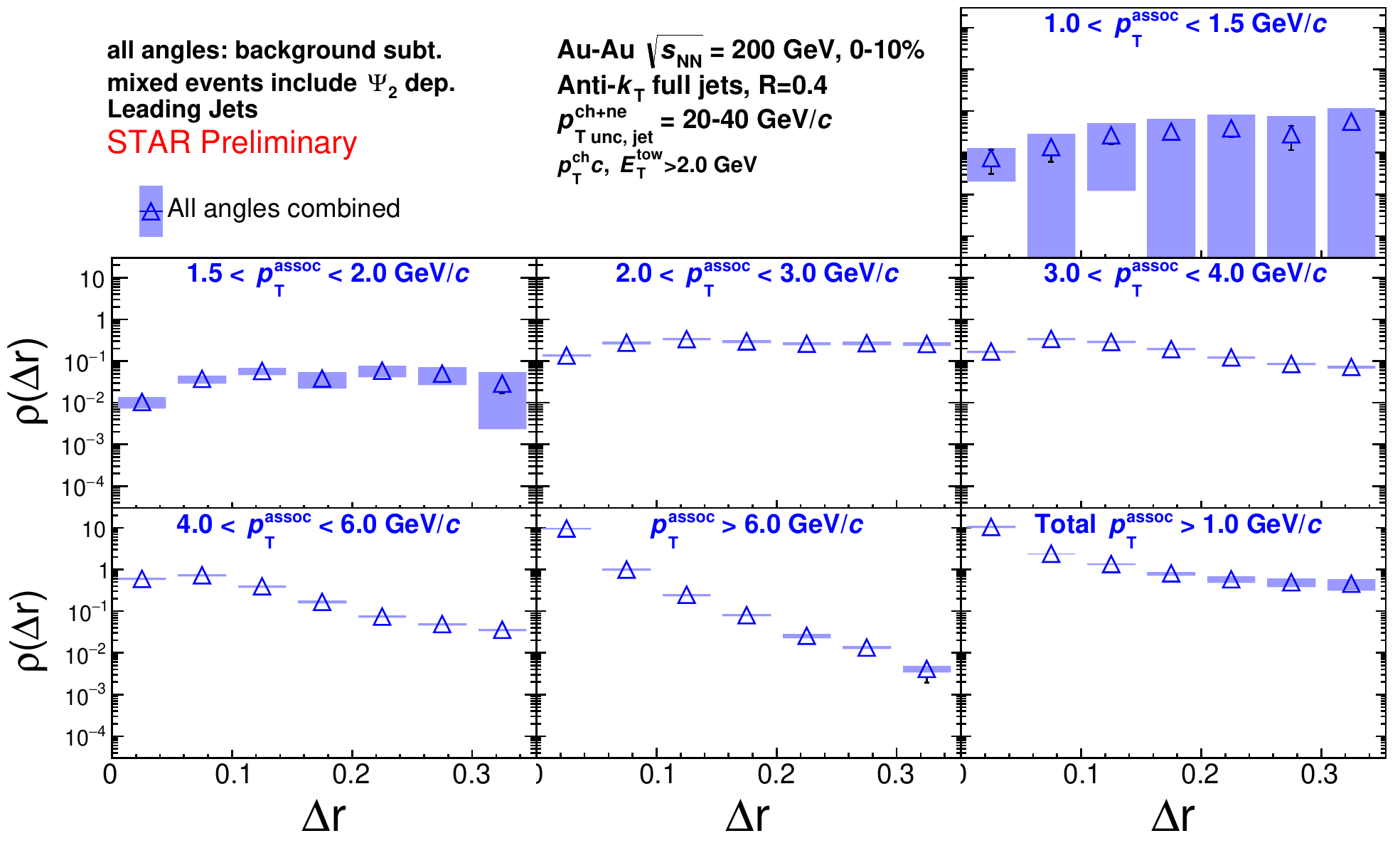}
	\caption{\label{fig:1} Differential jet shapes as a function of $r$ for 20-40 GeV/$c$ and $R=0.4$ full jets in 0-10\% central Au+Au collisions. Background contributions are estimated with the mixed-event technique and subtracted in each $p_{\mathrm{T}}^{\mathrm{assoc}}$ range.}
\end{figure}
Differential jet shape functions for 20-40 GeV/$c$ and $R=0.4$ full jets in 0-10\% central Au+Au collisions are shown in Fig.~\ref{fig:1}, where different panels represent different $p_{\mathrm{T,track}}$ (i.e., $p_{\mathrm{T}}^{\mathrm{assoc}}$) ranges. 
Statistical uncertainties and systematic uncertainties are represented with lines and colored boxes, respectively. 
Background contributions are estimated by associating measured jets with tracks in mixed events of the same event class, and subtracted accordingly. 
These results are corrected for tracking efficiency effects. 
While high-$p_{\mathrm{T}}$ tracks are more collimated to the jet axis (in part from the ``hard-core" jet selection and anti-$k_{\mathrm{T}}$ jet finding algorithm) compared to low-$p_{\mathrm{T}}$ tracks, the differential jet shape over all constituent momenta (bottom right panel) at this collision energy is observed to be broader than those at the LHC energies~\cite{Chatrchyan:2013kwa} with variations in kinematics and jet selection. 

\begin{figure}
	\centering
	\includegraphics[width=0.90\textwidth]{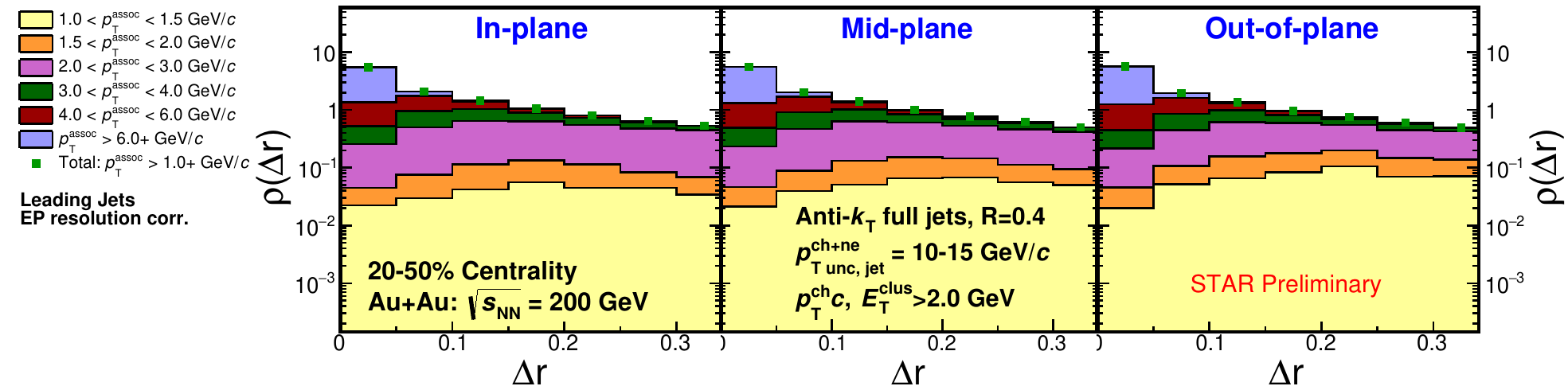}
	\caption{\label{fig:2} Event-plane dependent differential jet shapes for 10-15 GeV/$c$ and $R=0.4$ full jets in 20-50\% central Au+Au collisions. Jets are classified based on the azimuthal angle with respect to the second-order event plane: in-plane (left), mid-plane (mid), and out-of-plane (right). Different colored boxes represent different $p_{\mathrm{T}}^{\mathrm{assoc}}$ ranges.}
\end{figure}

In addition, the event-plane dependence of jet shapes is investigated for the first time in relativistic heavy-ion collisions. 
Jets are classified based on the azimuthal angle with respect to $\Psi_{\mathrm{EP}}$, measured following the same procedure in~\cite{Agakishiev:2014ada}, into in-plane ($0<\vert \varphi_{\mathrm{jet}} - \Psi_{\mathrm{EP}} \vert < \pi/6$), mid-plane ($\pi/6<\vert \varphi_{\mathrm{jet}} - \Psi_{\mathrm{EP}} \vert < \pi/3$), and out-of-plane ($\pi/3<\vert \varphi_{\mathrm{jet}} - \Psi_{\mathrm{EP}} \vert < \pi/2$) jets. 
Such classification may shed light on in-medium path-length dependent effects originating from the initial collision geometry.
Due to the almond shape of the initial geometry in 20-50\% central collisions, out-of-plane jets are expected to have a longer path length through the medium on average than in-plane jets. 
Figure~\ref{fig:2} shows event-plane dependent differential jet shape for 10-15 GeV/$c$ and $R=0.4$ full jets in 20-50\% central Au+Au collisions as an example. 
While high-$p_{\mathrm{T}}$ tracks are less sensitive to the azimuthal angle of jets with respect to $\Psi_{\mathrm{EP}}$, low-$p_{\mathrm{T}}$ tracks show a hint of dependence in their yields, which seem to be pushed toward farther distances in the out-of-plane direction relative to the in-plane direction.
The overall particle yield increases for out-of-plane jets in comparison to in-plane jets, particularly at larger distances from the jet axis. 
These observations may indicate larger in-medium path-length dependent effects in out-of-plane jets relative to in-plane jets. 

\section{Semi-inclusive jet fragmentation functions}
\label{sec:JetFF}
Jet fragmentation functions, $1/N_{\mathrm{jet}} \,\mathrm{d}N_{\mathrm{ch}}/\mathrm{d}z$, correspond to the distribution of constituent charged particle longitudinal momentum fraction with respect to the jet momentum normalized per jet, and have been previously reported by LHC collaborations~\cite{Aaboud:2017bzv, Chatrchyan:2012gw}. 
In order to have a proper handle on background jets, a semi-inclusive approach is used following similar procedures in~\cite{Adamczyk:2017yhe}. Charged jets in the recoil region ($3\pi/4 < \vert \varphi_{\mathrm{jet}} - \varphi_{\mathrm{trig}} \vert < 5\pi/4$) of a high momentum BEMC tower ($9.0 < E_{\mathrm{T}} < 30.0$ GeV)  are considered 
\begin{wrapfigure}{l}{0.48\textwidth}
	\centering
	\includegraphics[width=0.46 \textwidth]{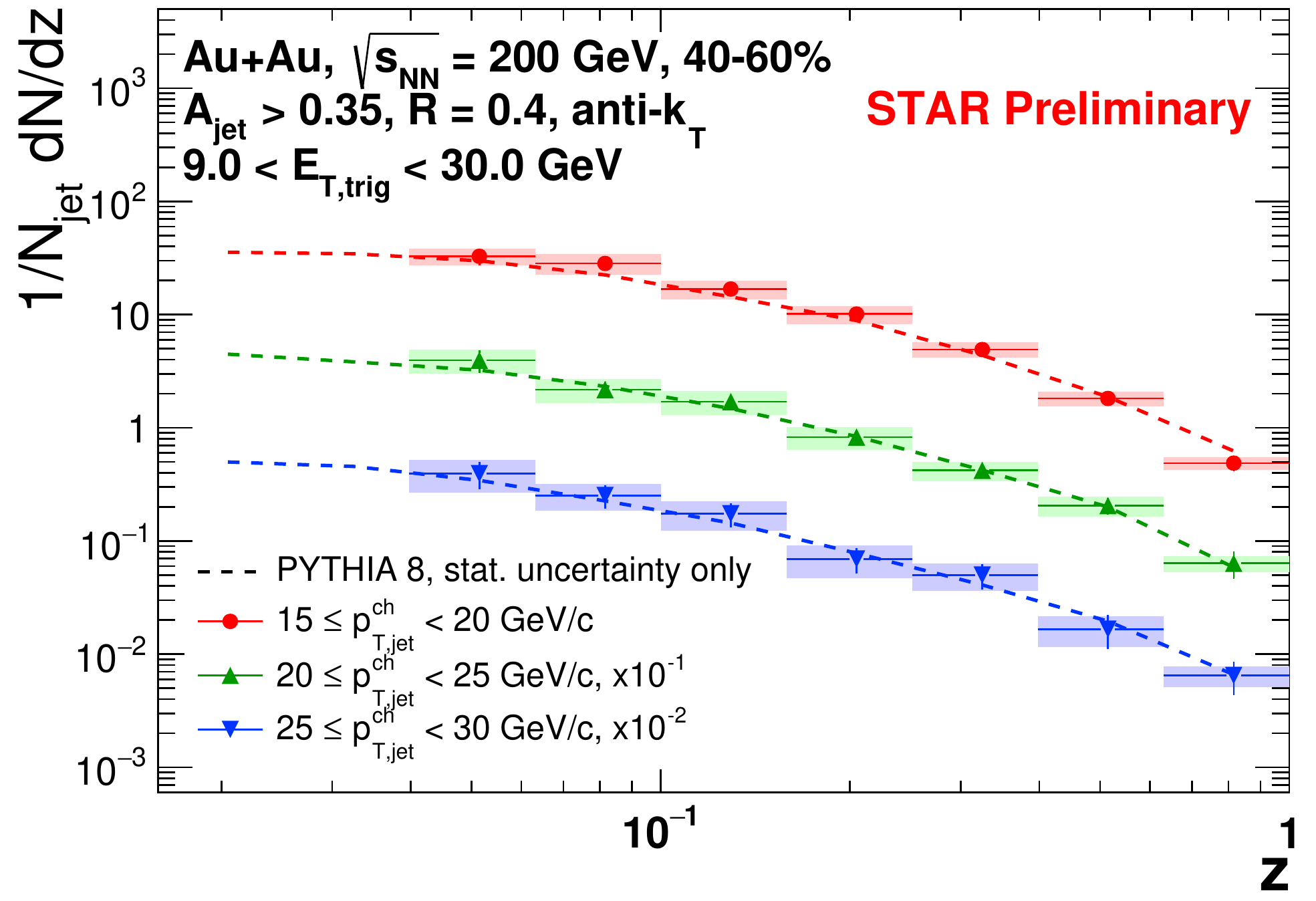}
	\caption{\label{fig:3} Semi-inclusive jet fragmentation functions in 40--60\% central Au+Au collisions for three $p_{\mathrm{T,jet}}$ ranges (closed markers), and PYTHIA 8 estimations for $pp$ collisions for the corresponding $p_{\mathrm{T,jet}}^{\mathrm{ch}}$ ranges (dashed lines). Data results are fully unfolded for detector effects and uncorrelated background effects.}
\end{wrapfigure}
in the measurement, and $z \equiv p_{\mathrm{T,track}}\,\mathrm{cos}(r)/p_{\mathrm{T,jet}}$ of charged tracks with $0.2 < p_{\mathrm{T,track}} < 30.0$ GeV/$c$ and $r < R$ is calculated by associating those charged tracks with the corresponding jet.

Uncorrelated components with respect to the trigger particle are removed independently in $N_{\mathrm{jet}}$ and $\mathrm{d}N_{\mathrm{ch}}/\mathrm{d}z$ via a mixed-event technique. 
For $N_{\mathrm{jet}}$, the number of jets in each $p_{\mathrm{T,jet}}$ bin, the same subtraction procedures as \cite{Adamczyk:2017yhe} are applied. 
For $\mathrm{d}N_{\mathrm{ch}}/\mathrm{d}z$, contributions from uncorrelated jets and uncorrelated particles in correlated jets are independently evaluated using mixed events, and subtracted accordingly. 
After such subtractions, $N_{\mathrm{jet}}$ and $\mathrm{d}N_{\mathrm{ch}}/\mathrm{d}z$ are independently unfolded via 1-dimensional and 2-dimensional Bayesian unfolding~\cite{Adye:2011gm}, respectively, for remaining uncorrelated background effects and instrumental effects in the fragmentation functions.  

\begin{figure}[h]
	\centering
	\includegraphics[width=0.84\textwidth]{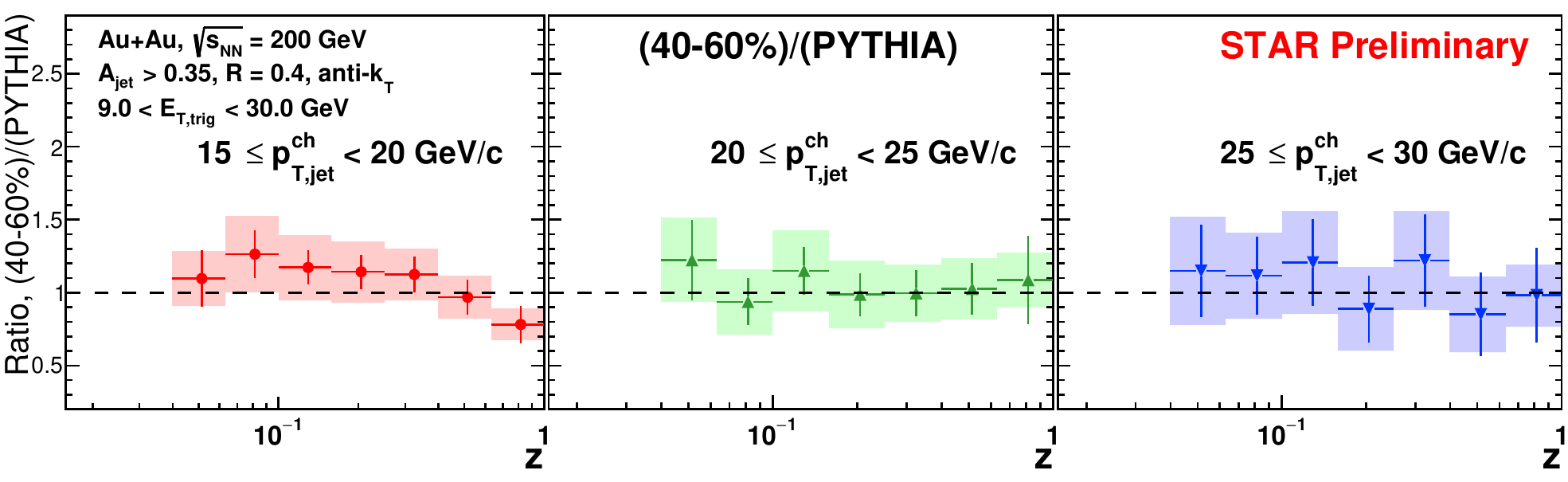}
	\caption{\label{fig:4} Ratios of jet fragmentation functions between 40--60\% central Au+Au collisions and PYTHIA 8 estimations for $pp$ collisions for three $p_{\mathrm{T,jet}}^{\mathrm{ch}}$ ranges.}
\end{figure}

Figure~\ref{fig:3} shows jet fragmentation functions for 40-60\% central Au+Au collisions and three $p_{\mathrm{T,jet}}$ ranges, along with PYTHIA 8 estimations for $pp$ collisions \cite{Adam:2020kug}, which was used in the recent STAR publication, for the corresponding $p_{\mathrm{T,jet}}$ ranges. 
Also, the ratios between 40-60\% central Au+Au collisions and PYTHIA 8 $pp$ estimations are shown in Fig.~\ref{fig:4}. 
It should be noted that only statistical uncertainties are included in PYTHIA 8 results. 
The ratios are observed to be consistent with unity within uncertainties throughout $z$ and $p_{\mathrm{T,jet}}$ ranges, which correspond to no significant modifications of jet fragmentation functions in 40-60\% central heavy-ion collisions at $\sqrt{s_{\mathrm{NN}}} = 200$ GeV. 
These results can be connected to various physics scenarios, for example, 1) tangential jet selection with a high-$p_{\mathrm{T}}$ trigger particle and recoil jet configuration, which causes no significant in-medium path length of the jet, 2) no significant jet-medium interactions in 40-60\% central heavy-ion collisions at $\sqrt{s_{\mathrm{NN}}}=200$ GeV collisions, 3) not enough path length of jets in the medium in 40-60\% centrality, and etc., and further studies are needed to differentiate them.  

\section{Outlook}
\label{sec:Outlook}
In these proceedings, preliminary differential jet shapes with the STAR experiment are reported. 
The full measurements include jet shapes with various jet finding parameter ($R$), those in central to peripheral collisions in addition to the $pp$ collisions, and a comparison among leading, sub-leading, and inclusive jets.
Meanwhile, semi-inclusive jet fragmentation functions are shown for 40-60\% central Au+Au collisions. 
Such measurements will be extended to the most central Au+Au collisions, and compared to those in $pp$ collisions as a vacuum reference. 
These results will be reported in the forthcoming publications, and should elucidate medium-induced modification of jet substructure at RHIC energies. 

\section*{Acknowledgement}
This work is supported by the US Department of Energy under award number DE-SC004168.











\end{document}